# How to measure research performance of single scientists? A proposal for an index based on scientific prizes: The Prize Winner Index (PWI)

Lutz Bornmann[1,2] and Robin Haunschild[2]

[1] *bornmann@gv.mpg.de*
Science Policy and Strategy Department, Administrative Headquarters of the Max Planck Society, Hofgartenstr. 8, 80539 Munich (Germany)

[2] *r.haunschild@fkf.mpg.de, l.bornmann@fkf.mpg.de*
Max Planck Institute for Solid State Research, Heisenbergstr. 1, 70569 Stuttgart (Germany)

**Abstract**
In this study, we propose a new index for measuring excellence in science which is based on collaborations (co-authorship distances) in science. The index is based on the Erdős number – a number that was introduced several years ago. We propose to focus with the new index on laureates of prestigious prizes in a certain field and to measure co-authorship distances between the laureates and other scientists. To exemplify and explain our proposal, we computed the proposed index in the field of quantitative science studies ($PWI_{PM}$). The Derek de Solla Price Memorial Award (Price Medal, PM) is awarded to outstanding scientists in the field. We tested the convergent validity of the $PWI_{PM}$. We were interested whether the indicator is related to an established bibliometric indicator: P(top 10%). The results show that the coefficients for the correlation between $PWI_{PM}$ and P(top 10%) are high (in cases when a sufficient number of papers have been considered for a reliable assessment of performance). Therefore, measured by an established indicator for research excellence, the new PWI indicator seems to be convergently valid and, therefore, might be a possible alternative for established (bibliometric) indicators – with a focus on prizes.

**Introduction**

Research without research evaluation is not imaginable; evaluations are an important and integral part of nearly all activities in science with the objective of research improvement and/or monitoring (Moed & Halevi, 2015). Research evaluation can be undertaken in a qualitative form with peer review processes; it is also very popular to evaluate research quantitatively using various indicators (Moed, 2017). An indicator is usually defined as a "measurable quantity that 'stands in' or substitutes for something less readily measurable and is presumed to associate with it without directly measuring it" (Wilsdon et al., 2015, p. 5). For example, citations may measure impact of research although impact may happen in a way that is not reflected in citations. The most frequently used indicators in research evaluation processes are bibliometric indicators (Szomszor et al., 2021). The results by Hammarfelt, Rushforth, and de Rijcke (2020) show, for example, that scientists prefer to assess candidates in peer review processes based on bibliometric indicators. One important (and understandable) reason of the popularity of the indicators is missing competence of the assessing scientists in the candidates' research. Besides simple bibliometric indicators such as the Journal Impact Factor, advanced indicators such as the disruption index or percentile-based field-normalized indicators are in use in today's research evaluation processes (Bornmann, 2019; Wang & Barabási, 2021).

With the h index, a very popular index for assessing the performance of single scientists has been introduced by Hirsch (2005), which combines publication output and citation impact in a single number. Besides the h index, many bibliometric indicators have been proposed for the use on the single researcher level in recent years, especially variants of the initial h index (Bornmann, Mutz, Hug, & Daniel, 2011). Some years ago, for example, Thomson Reuters (the former provider of the Web of Science, WoS, database; it is Clarivate now) introduced the Highly Cited Researchers database (see https://recognition.webofscience.com/awards/highly-cited) including researchers with the most papers belonging to the 1% most frequently-cited papers (Clarivate Analytics, 2021). Recently, Clarivate introduced Beamplots as an alternative to the popular h index to measure citation impact field-normalized and over the complete range of a scientist's publication years (Bornmann & Marx, 2014; Szomszor & Pendlebury, 2021). Although bibliometrics is frequently employed in the evaluation of single scientists, its use has often been criticized. According to Sunahara, Perc, and Ribeiro (2021), the use of bibliometrics "exerts enormous pressure on scholars (particularly on young scientists …) for publishing in large quantities, in prestigious journals, and developing highly cited research". In addition, many bibliometric indicators have been proposed in the past, and there is no consensus which indicator should be used (as standard). For example, performance can be measured size-dependently or size-independently. On the other side, the comparison of various h index variants by Bornmann et al. (2011) reveal that the variants measure performance similarly. Thus, there seems to be great redundancy between bibliometric indicators (on the single scientists' level).

In this study, we take up the critique on bibliometrics and leave the usual way of measuring performance by using citations. We propose an alternative measure that is based on collaborations (co-authorships) in science. The alternative is oriented towards the Erdős number – a proposal for measuring performance in mathematics. The Erdős number reflects the "collaborative distance" between the mathematician Paul Erdős (1913-1996) and other scientists (authors). Paul Erdős was an outstanding mathematician who published in collaboration with many other mathematicians. Here, we use the basic idea of the Erdős number – co-authorship distances – to propose a new performance metric – called Prize Winner Index (PWI) – for single scientists. We propose to focus with the new index on laureates of prestigious prizes in a certain discipline and to measure co-authorship distances between the laureates and other scientists.

To exemplify and explain our proposal, we computed the proposed index for the field of quantitative science. We selected this field, since we have both been active in this field for many years and are therefore in the position to interpret the empirical results. The Derek de Solla Price Memorial Award (Price Medal, PM) is awarded to outstanding scientists in quantitative science studies. We name the PWI for the field of quantitative science studies $PWI_{PM}$. We (RH) developed an R package (PWIR, Haunschild, 2022) that can be used to calculate the PWI for all authors in a dataset (e.g., from the Web of Science) with the names of the laureates as input by the user. The laureates of the PM are used as a default in the R package. Datasets from other databases could be converted into supported format(s).

**Prize Winner Index (PWI)**

Paul Erdős was one of the most important mathematicians, who received many prizes. He was enormously productive (in terms of his number of publications) and collaborated frequently with other and different scientists. According to Glänzel and Abdulhayoğlu (2018), it is not clear who initially proposed the Erdős number, but an explanation of it can already be found in Goffman (1969). A lot of information around the number and its calculation has been published on the website of the "Erdős Number Project" by Jerry Grossman, Professor of Mathematics, Emeritus at Oakland University.[1] The Erdős number is defined as "the shortest path connecting an author with Erdős in the complete co-author network created by Paul Erdős and his *n*-th order co-authors, which is iteratively generated by always adding the co-authors of co-authors, who are not already members of the network" (Glänzel & Abdulhayoğlu, 2018, p. 534). Based on this definition, Paul Erdős has an Erdős number of 0; his co-authors have an Erdős number of 1. All co-authors of Paul Erdős co-authors (i.e., co-authors who have not published directly with Paul Erdős) have an Erdős number of 2. Higher Erdős numbers follow for further co-authors of co-authors. Following Glänzel and Abdulhayoğlu (2018), the mean Erdős number of mathematicians is about 5.

---

[1] see https://sites.google.com/oakland.edu/grossman/home/the-erdoes-number-project

In this study, we propose – following the basic Erdős number definition – the PWI that counts co-authorships with prize winners (and their co-authors) in a certain discipline (usually). The PWI is defined as follows: The co-authorship distance from prize winners ($d_{pw}$, 0 for prize winners themselves, 1 for co-authors of prize winners, 2 for co-authors of co-authors of prize winners) is determined for each paper $p$ and each prize winner $w$. The formula $1/2^{d_{pw}}$ provides harmonic distant weights for each co-authorship distance value; e.g., 1, ½, ¼, and ⅛ are obtained for the co-authorship distances 0, 1, 2, and 3. The weights are summed up for each author over each paper and prize winner in the dataset. The PWI can be written as:

$$\text{PWI} = \sum_p \sum_w \frac{1}{2^{d_{pw}}}$$

The PWI is a size-dependent indicator. Scientists who have authored more papers are more likely to have a higher PWI. In addition to the PWI, the R package PWIR also provides the number of papers and the number of co-authors of each author in the dataset. This enables the user to easily compute relative PWI variants: PWI per published paper and PWI per co-authors. For matching as many synonyms of authors, only the last name and the first name (including other given names) initials are used by the R package PWIR, although this might cause homonyms to be merged together. Sensible author name disambiguation has to be carried out before by altering the author names (e.g., by adding further additional given names). Also, merging occurrences of synonyms requires preprocessing of the downloaded files.

To illustrate the calculation of the PWI, we used a publication set with two papers from the WoS. The exemplary prize winner is L. Waltman. To retrieve the two papers, we undertook a topic search for "gender differences in scientific careers" in the WoS. As of October 18, 2022, this search delivers two publications. The author lists of these two publications are shown in Table 1.

**Table 1. Publications with their author lists as retrieved from a topic search in Web of Science as of October 18, 2022**

| Publication Nº | List of authors |
|---|---|
| 1 | P. Mahlck |
| 2 | H. Boekhout, I. van der Weijden, L. Waltman |

The author P. Mahlck is not connected to L. Waltman. Thus, this author has a PWI value of zero. The author L. Waltman occurs once with a co-author distance of zero ($d_{pw} = 0$): PWI = $1 \cdot 1/2^0 = 1$. The authors H. Boekhout and I. van der Weijden occur once with a co-author distance of one ($d_{pw} = 1$): PWI = $1 \cdot 1/2^1 = .5$. The corresponding output of the function PWI from R package PWIR when the WoS download as described above is provided as input is shown in Table 2. The output includes the number of papers and the number of co-authors besides the PWI.

**Table 2. Output of the function PWI from the R package PWIR when the Web of Science download as described above is provided as input**

| Author | PWI | Number of papers | Number of co-authors |
|---|---|---|---|
| WALTMAN L | 1 | 1 | 2 |
| BOEKHOUT H | .5 | 1 | 2 |
| VAN DER WEIJDEN I | .5 | 1 | 2 |
| MAHLCK P | 0 | 1 | 0 |

The PWI is based on two premises: (1) In every discipline, "scientific elites" exist that can be identified by (prestigious) prizes. According to Zuckerman (1977), scientific elites "are worthy of our attention not merely because they have prestige and influence in science, but because their collective contributions have made a difference in the advance of scientific knowledge" (as cited in Li, Yin, Fortunato, & Wang, 2020). In her study, Zuckerman (1977) identified scientific elites by the Nobel prize, i.e., the elite received this prestigious prize. For Tijssen (2020), "Nobel prizes are often considered, especially by the general public, to be an ultimate accolade of international excellence" (p. 59). (2) Scientists who collaborate with the elite or with scientists in the narrow collaboration network of the elite do research on a high research quality level.

In the following, we would like to explain our rationale to base the PWI on the two premises in more detail: (Ad 1) In the reward and incentives system of science, prizes have an important value. For Ma and Uzzi (2018), prizes "identify top scientific achievements … identify successful role models who inspire achievements once thought to be impossible … and act as signals of scientific credibility … Prizes may also forecast the direction of future scientific investments. Prizewinning papers are cited in patents faster than similarly cited, non-prizewinning papers … and often include prizewinners with direct or indirect capital (e.g., Howard Hughes Medical Research Award) that stimulates research" (p. 12608). The empirical results by the authors based on bibliometric data show that "prizes are more concentrated within a relatively small scientific elite" (p. 12608). Prizewinning topics do not only have significant increases in growth (productivity, citation impact, and new entrants) (Jin, Ma, & Uzzi, 2021), scientific prizes are also meaningful events in the career of scientists leading to changes in publication practices (Liu, Yu, Chen, & Huang, 2018) and the receipt of additional prizes (Chan, Mixon, & Torgler, 2018; Zheng & Liu, 2015).

(Ad 2) The second premise for the new index refers to the rational of basing the PWI on collaborations with elite scientists and authors in their collaboration network. Literature overviews on collaborations in science have been published by Bozeman, Fay, and Slade (2013) and Katz and Martin (1997). According to Zeng, Fan, Di, Wang, and Havlin (2021) "teamwork is becoming increasingly common in recent modern science". The mean number of co-authors per paper has nearly doubled since the 1950s (Zeng et al., 2021), and the number of solo-authored papers dramatically decreased (Wang & Barabási, 2021). The studies by Cimenler, Reeves, and Skvoretz (2014) and Milojevic, Radicchi, and Walsh (2018) show that career success and growth are related to scientists' number of collaborators and prestige of advisors. Elite scientists tend to cooperate with other strong scientists (or promising young scientists) and have a stimulating effect on their research environment (Wang & Barabási, 2021). The significant influence of strong collaborations on scientific performance has been denoted as apostle effect (Gallotti & De Domenico, 2019). This effect has been demonstrated especially for young scientists (Li, Aste, Caccioli, & Livan, 2019).

## Methods

*Price Medal (PM)*

In this study, we computed the PWI for the field of quantitative science studies. Scientists active in science studies are organized in the International Society for Scientometrics and Informetrics (ISSI). The community focuses on quantitative approaches to the study of science that include informetrics, scientometrics, and webometrics. ISSI provides four awards to acknowledge exemplary achievements in the community. The highest award is the PM. It is the only award in this field that refers to lifetime a achievement. Since 1984, the PM is awarded to outstanding scientists in quantitative science studies. The other awards honor single papers, students, or doctoral students. The PM has been named after the scientist Derek de Solla Price, who can be regarded as "one of the founders of scientometrics. He published extensively in the 1960s and 1970s … laying the foundations for the newly emerging field of quantitative science studies" (Wyatt, Milojevic, Park, & Leydesdorff, 2016, p. 88).

On the ISSI homepage, the PM is explained as follows: "The Price Medal was conceived and launched by Tibor Braun, founder and former Editor-in-Chief of the international journal *Scientometrics*, and is periodically awarded by the journal to scientists with outstanding contributions to the fields of quantitative studies of science. The journal *Scientometrics* is an international journal for all quantitative aspects of the science of science, communication in science and science policy co-published by Akadémiai Kiadó, Budapest, and Springer, Dordrecht. The first medal was awarded to Eugene Garfield in 1984" (https://www.issi-society.org/awards/derek-de-solla-price-memorial-medal, accessed at November 8, 2022). Eugene Garfield can be seen as one of the 'fathers' of scientometrics who developed the first citation index (Garfield, 1955).

Because of the great importance of Eugene Garfield for the field of scientometrics and his intensive publication and collaboration activities over many years, Glänzel and Abdulhayoğlu (2018) calculated the Erdős number for scientometrics based on the papers published by Eugene Garfield.

*Datasets*

In this study, we used three journal sets that reflect the field of science studies in a closer or wider perspective. We do not focus on only one set, since the field can be differently defined, and we are interested how robust our empirical results are. Robust results would be reflected in similar results based on different datasets.

*Core journals*: Although the community of a given field publishes its results mostly in specific journals, there are usually core journals and more peripheral journals. In the field of science of science studies, *Scientometrics*, *Journal of Informetrics* (JOI), and *Quantitative Science Studies* (QSS) can be considered as core journals. *Scientometrics* is not only the eldest journal in the field with the most annual papers but also the journal that awards the PM. QSS is the official journal of the ISSI. JOI can be indicated as core journal, since QSS resulted from JOI: Some years ago, the chief and deputy editors and the editorial board of JOI decided to leave JOI and to switch to the newly established QSS. We downloaded 8,393 records from the WoS on July 21, 2022, of the aforementioned three sources.

*iMetrics*: Leydesdorff, Bornmann, Marx, and Milojevic (2014) and Maltseva and Batagelj (2020) used the term iMetrics to denote the field of scientometrics based on paper and journal selections. The iMetrics set is mostly identical with our core journals set that additionally includes, however, the *Journal of the American Society for Information Science and Technology* (the alternative title *Journal of the Association for Information Science and Technology* was also considered). We downloaded 12,473 records from the WoS between July, 21 and 25, 2022, of the aforementioned four sources.

*Complete set*: The complete set includes papers from journals and proceedings that publish papers from the science studies field. The proceedings of the ISSI and the publications in the following eight journals seem reasonable to us to cover the wider field of science of science studies in that the PM is centered: *Journal of Data and Information Science*, *Journal of Information Science*, *Journal of Informetrics*, *Journal of the American Society for Information Science and Technology* (the alternative title *Journal of the Association for Information Science and Technology* was also considered), *Profesional de la Informacion*, *Quantitative Science Studies*, *Research Evaluation*, and *Scientometrics*. We downloaded 19,166 records from the WoS between July 21 and 25, 2022, of the aforementioned nine sources.

We used the datasets and the PWI function to produce lists of authors with $PWI_{PM}$ values. The initial lists showed the need for a proper author name disambiguation. For example, the entry VAN RAAN AFJ belongs to the same person as the entries VANRAAN AFJ and VAN RAAN A, respectively. Thus, the $PWI_{PM}$ values had to be recalculated when author names were merged. The function PWI provides an option "method=0" to just print the author names, number of papers, and number of co-authors for checking of author name variants. These lists can be used to identify variants of author names and to disambiguate the names in the WoS input files. Afterwards, the function PWI can be run without the option "method=0" for computing the "correct" $PWI_{PM}$ values.

For this study, we identified and merged the names of the PM awardees in the three datasets and received the following numbers of authors: core journals = 9,802; iMetrics = 14,228; complete set = 20,903.

*Statistics*

We analyzed the relationship between $PWI_{PM}$ values and number of papers published, number of co-authors, and the status of an author of being a PM laureate or not. Since we are interested in how much each of the three variables contributes uniquely to the $PWI_{PM}$ values of the authors, we performed a robust multiple regression analysis with $PWI_{PM}$ values as dependent and number of papers published, number of co-authors, and status of an author of being a PM laureate or not as independent variables. We decided to compute a robust regression (Acock, 2018), since the $PWI_{PM}$ values are skewed distributed (see Figure 1). We used the program Stata for the statistical analyses of this study (StataCorp., 2021).

*Definition of P(top 10%)*

For the validation of the $PWI_{PM}$, we correlated the indicator values with P(top 10%). We used P(top 10%) values from our WoS custom database. The P(top 10%) values were calculated according to the procedure proposed by Waltman and Schreiber (2013). This ensures that exactly 10% of the papers are top 10% papers, however, at the expense that some papers are only partial top 10% papers and obtain a fractional P(top 10%) value.

**Results**

The full list of $PWI_{PM}$ values for authors in the field of science studies based on the three datasets is available at: https://ivs.fkf.mpg.de/PWI/PWI-data.xlsx. Figure 1 shows cumulative probability plots of $PWI_{PM}$ values separated for the status of an author (of being an awardee of the PM or not) based on the three datasets. The results for the datasets are very similar. First, the plots reveal that the values are very skewed distributed: There are only a few authors with high $PWI_{PM}$ values. Second, the plots show that there are great differences between awardees and non-awardees: The few authors with high $PWI_{PM}$ values are mostly awardees.

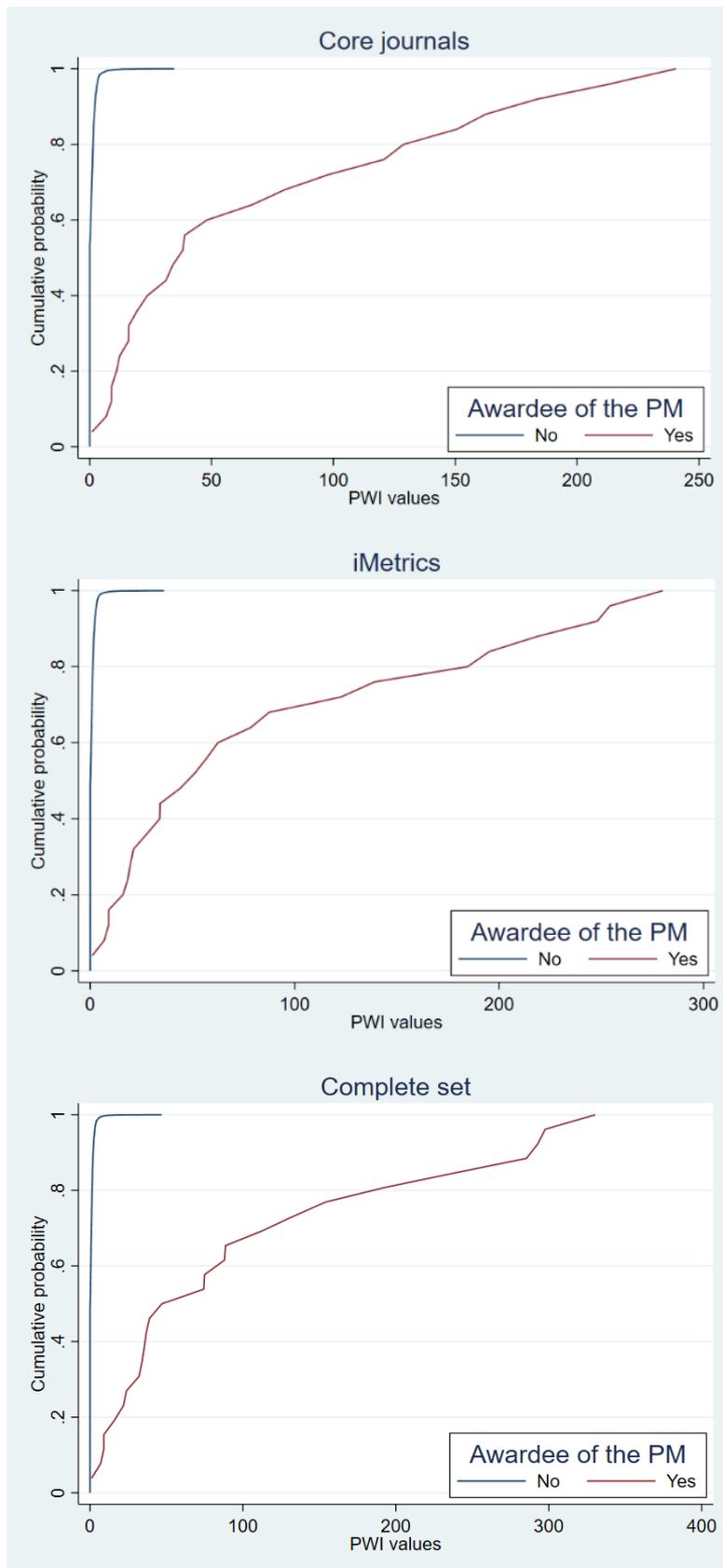

Figure 1. Cumulative probability plots of PWI$_{PM}$ values separated for the status of an author (of being an awardee of the PM or not) based on three datasets

*Authors with the highest PWI$_{PM}$ values*

The top-20 authors of the scientometrics sources (using the three datasets) ordered by PWI$_{PM}$ values in descending order are shown in Table 3. In accordance with the results in Figure 1, the results in the table reveal that awardees of the PM are in the best position to receive high PWI$_{PM}$ values: Nearly all authors in the table are awardees who have been active in the field of science studies for many years. The high number of awardees among the top authors can be expected and was conceived, since the PWI$_{PM}$ is strongly oriented towards the PM. We would like to highlight in Table 3 the exceptional achievement of Peter Vinkler (Price Medalist in 2009), who published all papers alone (without any contribution of co-authors). There are five scientists in Table 3 who did not receive the PM: Hans-Dieter Daniel, Nees Jan van Eck, Robin Haunschild, Bart Thijs, and Lin Zhang. The five scientists have collaborated extensively with one Price Medalist each: Wolfgang Glänzel (Bart Thijs, Lin Zhang), Lutz Bornmann (Hans-Dieter Daniel, Robin Haunschild), and Ludo Waltman (Nees Jan van Eck).

Table 3. Output of the function PWI$_{PM}$ from the R package PWIR when the Web of Science downloads as described above is provided as input

| Author | PWI$_{PM}$ | Number of papers | Number of co-authors | Price Medalist |
|---|---|---|---|---|
| *Core journals* | | | | |
| GLANZEL W | 240.50 | 187 | 101 | Yes |
| BORNMANN L | 213.53 | 187 | 81 | Yes |
| SCHUBERT A | 183.75 | 129 | 28 | Yes |
| LEYDESDORFF L | 162.56 | 138 | 66 | Yes |
| ROUSSEAU R | 150.63 | 133 | 94 | Yes |
| THELWALL M | 128.78 | 125 | 67 | Yes |
| BRAUN T | 120.75 | 74 | 16 | Yes |
| EGGHE L | 97.94 | 85 | 9 | Yes |
| MOED HF | 80.00 | 66 | 60 | Yes |
| VAN RAAN AFJ | 66.53 | 56 | 25 | Yes |
| WALTMAN L | 48.16 | 42 | 31 | Yes |
| VINKLER P | 39.00 | 39 | 0 | Yes |
| BAR-ILAN J | 38.16 | 33 | 19 | Yes |
| THIJS B | 34.69 | 41 | 28 | No |
| MORAVCSIK MJ | 34.00 | 34 | 5 | Yes |
| PERSSON O | 31.28 | 23 | 18 | Yes |
| HAUNSCHILD R | 25.53 | 44 | 27 | No |
| INGWERSEN P | 23.66 | 20 | 22 | Yes |
| DANIEL HD | 22.77 | 36 | 15 | No |
| ZHANG L | 21.06 | 40 | 63 | No |
| *iMetrics* | | | | |
| BORNMANN L | 280.05 | 246 | 93 | Yes |
| LEYDESDORFF L | 254.20 | 221 | 104 | Yes |
| GLANZEL W | 248.03 | 193 | 104 | Yes |
| THELWALL M | 218.92 | 214 | 95 | Yes |
| ROUSSEAU R | 195.28 | 171 | 107 | Yes |
| SCHUBERT A | 184.52 | 129 | 28 | Yes |
| EGGHE L | 139.09 | 122 | 12 | Yes |
| BRAUN T | 122.52 | 75 | 17 | Yes |
| MOED HF | 87.52 | 73 | 63 | Yes |
| VAN RAAN AFJ | 78.54 | 67 | 30 | Yes |
| WALTMAN L | 62.42 | 55 | 38 | Yes |
| BAR-ILAN J | 57.05 | 51 | 32 | Yes |
| CRONIN B | 51.16 | 47 | 21 | Yes |
| VINKLER P | 44.00 | 44 | 0 | Yes |
| THIJS B | 36.08 | 42 | 28 | No |
| PERSSON O | 34.17 | 26 | 20 | Yes |
| MORAVCSIK MJ | 34.00 | 34 | 5 | Yes |
| DANIEL HD | 29.27 | 48 | 16 | No |
| INGWERSEN P | 27.69 | 24 | 25 | Yes |
| HAUNSCHILD R | 27.55 | 48 | 28 | No |
| *Complete set* | | | | |
| BORNMANN L | 330.41 | 290 | 108 | Yes |
| LEYDESDORFF L | 297.69 | 258 | 125 | Yes |
| GLÄNZEL W | 292.81 | 234 | 118 | Yes |
| THELWALL M | 285.41 | 279 | 109 | Yes |
| ROUSSEAU R | 238.63 | 211 | 116 | Yes |
| SCHUBERT A | 192.91 | 134 | 45 | Yes |
| EGGHE L | 154.13 | 135 | 14 | Yes |
| BRAUN T | 132.66 | 82 | 18 | Yes |
| MOED HF | 112.31 | 97 | 70 | Yes |
| WALTMAN L | 88.66 | 80 | 53 | Yes |
| VAN RAAN AFJ | 87.94 | 76 | 36 | Yes |
| BAR-ILAN J | 74.91 | 65 | 37 | Yes |
| CRONIN B | 74.41 | 69 | 25 | Yes |
| VINKLER P | 47.00 | 47 | 0 | Yes |
| THIJS B | 46.88 | 62 | 34 | No |
| INGWERSEN P | 38.91 | 35 | 33 | Yes |
| PERSSON O | 36.78 | 28 | 22 | Yes |
| DANIEL HD | 35.70 | 61 | 20 | No |
| HAUNSCHILD R | 35.59 | 62 | 31 | No |
| GARFIELD E | 35.44 | 32 | 9 | Yes |

The researchers in Table 3 are well-known in the field of scientometrics. They can be found as chief editors of important journals in the field. For example, Ludo Waltman is chief editor of QSS and the former editor of JOI. He took over the position at JOI from Leo Egghe, who was the first editor of JOI. Wolfgang Glänzel has been chief editor of *Scientometrics* for many years; Tibor Braun is the founder and former chief editor of this journal. All researchers in Table 3 are current or former members of editorial boards of scientometric journals. Researchers in Table 3 can also be found in the Highly Cited Researchers database (in social sciences), which is published annually by Clarivate (Clarivate Analytics, 2021, and earlier by Thomson Reuters) such as Loet Leydesdorff, Mike Thelwall, and Ludo Waltman. In 2019, Springer published the Handbook of Science and Technology Indicators, which includes "state-of-the-art descriptions of the wide variety of indicators and methods used for research and innovation assessment" (Glänzel, Moed, Schmoch, & Thelwall, 2019). Three of the four editors belong to the twenty researchers in the field with the highest $PWI_{PM}$ values.

The results in Table 3 have been produced by the R package PWIR. This package does not only compute individual $PWI_{PM}$ values but also the number of papers and number of co-authors. The results in the table show that the PWI$_{PM}$ values probably correlate with the number of papers and number of co-authors. The values seem to depend on these numbers. Since we are interested in the variables that explain variation in PWI$_{PM}$ values, we performed several regression models. The results are presented in the next section.

*Relationships between PWI$_{PM}$, number of papers, and number of co-authors*

The results in Table 3 of the authors with high PWI$_{PM}$ values indicate that researchers with many papers in the field and in collaboration with other researchers in the community are in the best position to publish (frequently) with awardees of the PM or are awardees themselves. In this section, therefore, we would like to test the dependence of the PWI$_{PM}$ values on the number of papers published, the number of co-authors, and the status of being an awardee or not.

**Table 4. Multiple robust regression analyses predicting PWI$_{PM}$ values from number of papers, number of co-authors, and being an awardee of the PM (or not)**

| Independent variables | Coefficient | 95% Confidence interval | Beta | Semipartial $R^2$ | $R^2$ | F | Number of authors |
|---|---|---|---|---|---|---|---|
| *Core journals* | | | | | .76 | 46.54*** | 9,801 |
| Constant | -.32** | [-.55, -.09] | | | | | |
| Number of papers | .71*** | [.48, .95] | .72 | .22*** | | | |
| Number of co-authors | -.10* | [-.19, -.01] | -.10 | .01*** | | | |
| Awardee (yes) | 31.94*** | [19.85, 44.03] | .31 | .07*** | | | |
| *iMetrics* | | | | | .78 | 49.14*** | 14,227 |
| Constant | -.23* | [-.43, -.02] | | | | | |
| Number of papers | .72*** | [.52, .92] | .75 | .23*** | | | |
| Number of co-authors | -.13** | [-.21, -.05] | -.13 | .01*** | | | |
| Awardee (yes) | 39.62*** | [25.33, 53.91] | .31 | .07*** | | | |
| *Complete set* | | | | | .75 | 45.37*** | 20,902 |
| Constant | -.07 | [-.25, .10] | | | | | |
| Number of papers | .66*** | [.46, 85] | .74 | .21*** | | | |
| Number of co-authors | -.15** | [-.23, -.06] | -.16 | .01*** | | | |
| Awardee (yes) | 50.57*** | [33.87, 67.26] | .34 | .08*** | | | |

Notes. * $p<.05$, ** $p<.01$, *** $p<.001$

The results of the regression analyses based on the three datasets are shown in Table 4. The results include not only the coefficients from the analyses, but also the $R^2$ and semipartial $R^2$. Whereas the $R^2$ reveal how much variation in the dependent variable is explained by the independent variables, the semipartial $R^2$ reveal the increment in $R^2$ by an independent variable. The $R^2$ values of the three models show that around 76% of variation in PWI$_{PM}$ values can be explained by the three independent variables. The results of the three overall $F$ tests indicate that there is a statistically significant relationship between $PWI_{PM}$ values and the three variables. Both results ($R^2$ values and $F$ tests) point out that the three independent variables are closely related to the $PWI_{PM}$ values.

The constants (or intercepts) in Table 4 indicate $PWI_{PM}$ values when an author is not an awardee and has published no paper (without any co-authors). Since this constellation does not exist, the constant is negative in all cases and not very informative. The coefficients can be used yet to calculate estimated $PWI_{PM}$ values: For example, the $PWI_{PM}$ value that is estimated based on the number of papers and co-authors using the core journals dataset is -.32 + .71 (number of papers) + (-.10) (number of co-authors). Thus, the estimated $PWI_{PM}$ value for an author (who is not an awardee) of ten papers with one co-author is -.32 + (.71 * 10) + (-.10 * 1) = 7.32. The status of being a Price Medalist leads – as expected – to a significant increase in the estimated $PWI_{PM}$ value (for an author with only one paper and co-author): -.31 + (.75 * 1) + (-.10 * 1) + (31.94 * 1) = 39.26. The estimated values may deviate stronger from observed values, since the distribution of PWI values is very skewed.

The results of the regression models in Table 4 based on the three datasets are very similar. The most important variable with respect to the $PWI_{PM}$ values seems to be the number of papers. The beta coefficient in the table is a measure of effect size that can be interpreted like a correlation with .1 being weak, .3 being moderate, and .5 being strong (Cohen, 1988). The beta coefficients of the number of papers are strong (the beta coefficients of the other variables are on a moderate or weak level). The semipartial $R^2$ values point to the same direction: The increments in $R^2$ are significantly higher for the number of papers (about .22) compared to both other independent variables (less than .1).

*Validation of the $PWI_{PM}$*

In this paper, we introduce with the PWI a new metric for research evaluation purposes that might be an alternative and an addition to establish metrics. With the introduction of such a new metric, one wonders how well it functions and whether it is able to measure what it proposes to measure (Kreiman & Maunsell, 2011; Moed, 2016). In the introduction section, we argue that the PWI uses information on scientific prizes to focus on measuring excellence (elite) in a certain field. In the following, we demonstrate exemplarily that authors with high $PWI_{PM}$ values are members of editorial boards and chief editors. Furthermore, authors of the most important handbook in scientometrics can be found among the authors with high $PWI_{PM}$ values. We also checked whether authors with high $PWI_{PM}$ values can be found in Clarivate's Highly Cited Researchers database.

In this section, we want to connect to these selective observations of validity by calculating the number of excellent papers – P(top 10%) – for all authors in our three datasets. We calculated the sum of the P(top 10%) values of the articles and reviews in our three datasets that were published before 2020 to allow for a citation window of at least three years. By correlating this number with $PWI_{PM}$ values, we assess the convergent validity of the new indicator (National Research Council, 2012). The convergent validity of an indicator is given if its correlation with a theoretically similar measure is "high" (Chen, Shao, & Fan, 2021; Jirschitzka, Oeberst, Göllner, & Cress, 2017). In statistical terms, "high" would mean that the correlation is at least .5 (Cumming, 2012). Table 5 shows the result of the correlation analyses based on the three datasets. The table presents the correlation coefficients and the number of authors included in the analyses. We calculated the coefficients with different thresholds for the number of papers per author, since we observed that the coefficients depend on the number of papers published by an author.

**Table 5. Spearman rank correlation coefficients between PWI$_{PM}$ values and P(top 10%) values**

| Threshold for the number of papers per author | Coefficient | Number of authors |
|---|---|---|
| *Core journals* | | |
| 1 | .24*** | 7,594 |
| 10 | .47*** | 191 |
| 20 | .63*** | 63 |
| 30 | .73*** | 28 |
| 40 | .80*** | 15 |
| 50 | .73** | 12 |
| *iMetrics* | | |
| 1 | .19*** | 11,585 |
| 10 | .47*** | 277 |
| 20 | .58*** | 97 |
| 30 | .66*** | 47 |
| 40 | .76*** | 27 |
| 50 | .86*** | 16 |
| *Complete set* | | |
| 1 | .18*** | 17,036 |
| 10 | .47*** | 431 |
| 20 | .52*** | 168 |
| 30 | .62*** | 77 |
| 40 | .65*** | 46 |
| 50 | .80*** | 28 |

Note. ** *p*<.01, *** *p*<.001

As the results in Table 5 reveal, the convergent validity of the PWI$_{PM}$ values is only then given when authors with at least ten papers per author are required. This limitation in the results seems reasonable since a reliable measurement of quality (performance) is only possible with a sufficient database (i.e., a sufficient number of papers per author). For Lehmann, Jackson, and Lautrup (2008), it is only possible "to draw reliable conclusions regarding an author's citation record on the basis of approximately 50 papers" (p. 384). Similar thresholds for the number of papers can be found in Glänzel and Moed (2013).

**Discussion**

According to Moed (2018) "government funding of scientific research is increasingly based upon performance criteria". For that, research evaluation needs indicators that are able to measure what is in the focus of a specific evaluation (Bornmann & Marewski, 2019). Many evaluation processes are dominated by the strive for excellence: "Excellence is omnipresent in the research ecosystem" (Jong, Franssen, & Pinfield, 2021). Only the elite among possible candidates should be at the top of the list in research evaluation processes. The most popular indicators for measuring excellence in science are bibliometric indicators. The Highly Cited Researchers database published by Clarivate Analytics (2021) reflects this popularity very well. In this study, we propose a new indicator which is not based on bibliometric data, but on scientific prizes. With the focus on prizes, the indicator follows a common feature behind many processes in science: "an amazingly steady tendency to the concentration of items on a relatively small stratum of sources" (de Bellis, 2009, p. xxiv). Since scientific prizes are – as a rule – rare events, only happening for exceptional scientists (Zuckerman, 1977), we used prizes as a starting point for the identification of scientific elites following the basic Erdős number definition. The PWI counts co-authorships with prize winners (and their co-authors) in a certain field; it is based on harmonic distant weights for each co-authorship distance value. Since we are interested in the narrow environments of prize winners, the PWI does not only focus on the prize winners themselves but also on scientists who have collaborated with prize winners and their co-authors. The PWI can be applied in various research evaluation processes such as funding of disruptive research, recruitment of exceptional researchers for professorships, and identification of reviewers for specific programs that support breakthrough research.

To exemplify the calculation and use of the PWI, we calculated the index for our own field (quantitative science studies) using three datasets that are linked to the PM. $PWI_{PM}$ values reflect the performance of authors who have published in the field of science studies. We observed that the authors with the highest $PWI_{PM}$ values are well-known scientists in the field being members of editorial boards and chief editors (among other things). We tested the dependence of $PWI_{PM}$ values on the number of papers, the number of co-authors, and the status of an author of being an awardee of the PM. The results of regression models show that the number of papers is the most important factor for receiving high $PWI_{PM}$ values, i.e., the index is size-dependent. For the calculation of a size-independent variant of the indicator, the $PWI_{PM}$ values can be divided by the number of papers or the number of years since publishing the first paper (which would be an age-normalized variant of the index). The regression models also reveal that another important factor for high $PWI_{PM}$ values is the status of being an awardee of the PM. This result was expected, because it reflects the design of the indicator.

The most important step of the empirical analyses in this study was the validation of the $PWI_{PM}$: Only validated indicators should be considered in the research evaluation practice. We tested the convergent validity of the $PWI_{PM}$. We were interested whether the indicator is related to an established bibliometric indicator: P(top 10%). The results show that the coefficients for the correlation between $PWI_{PM}$ and P(top 10%) are high (in cases when a sufficient number of papers have been considered for a reliable assessment of performance). Therefore, measured by an established indicator for research excellence (Bornmann, de Moya Anegón, & Leydesdorff, 2012), the new PWI indicator seems to be convergently valid. Since this conclusion is based on only one indicator to measure the convergent validity of the PWI, future studies should use other indicators to confirm our results (or not). Future studies could also try to (in-)validate the PWI in other fields using other prizes. With the developed R package (PWIR, Haunschild, 2022), PWI values can be computed for all downloads from the WoS database.

In principle, we see several areas for a possible optimization of the PWI. Future studies might focus on these areas with empirical investigations: (1) We already mentioned that the PWI could be age-normalized by dividing the value by the number of years since publishing the first paper. (2) Another area concerns the point in time of the prize award. In the current definition of the PWI, it is not considered when the prize was awarded. Authors are classified as prize winners independent of the point in time of winning the prize. One could argue that authors are prize winners only from that point in time when they received the prize. The consideration of the point in time leads to a more complicated calculation of the PWI and may change PWI values. (3) The index assumes voluntary collaborations, i.e., collaborations that have been stricken up based on research quality considerations. However, many collaborations have other/additional reasons. For example, we identified some authors from science studies with high $PWI_{PM}$ values with employee relationships. On one side, one may argue that these relationships result from quality considerations. On the other side, however, the relationships make author collaborations more likely than for two authors without any employee relationship. Thus, it might be necessary to consider employee relationships at least in the interpretation of the results. (4) Since price winners dominate the PWI results, an alternative PWI may leave out price winners in the results. This alternative would measure only collaborations with price winners and their co-authors.

As with all other indicators, the PWI is imperfect or biased. It focuses on a certain part of scientific activity, and therefore, it measures scientific performance in a biased way. According to Moed (2017), the awareness that "all performance indicators are 'partial' or 'imperfect' – influenced as they may be not only by the various aspects of performance but also by other factors that have little to do with performance – is as old as the use of performance indicators itself. Indicators may be imperfect or biased, but in the application of such indicators this is not seldom forgotten" (p. 6). From the awareness that indicators are imperfect or biased, one should follow that "metrics should support, not supplant, expert judgement. Peer review is not perfect, but it is the best form of academic governance we have, and it should remain the main basis by which to assess research papers, proposals and individuals" (Wilsdon, 2015). The royal road of research evaluation lies in "the intelligent combination of metrics and peer review" (Moed & Halevi, 2015).

**Acknowledgments**

The bibliographic data used in this paper are from the online version of the WoS provided by Clarivate. The P(top 10%) data used in this paper are from a WoS custom database of the Max Planck Society (MPG) developed and maintained in cooperation with the Max Planck Digital Library (MPDL, Munich) and derived from the Science Citation Index Expanded (SCI-E), Social Sciences Citation Index (SSCI), Arts and Humanities Citation Index (AHCI), Conference Proceedings Citation Index-Science (CPCI-S), and Conference Proceedings Citation Index-Social Science & Humanities (CPCI-SSH) provided by Clarivate via the "Kompetenznetzwerk Bibliometrie" (see https://bibliometrie.info/en/about-kb/) funded by BMBF (grant 16WIK2101A).